# Empêchée, contrainte ou souhaitée : trois visages de la reconversion au prisme des catégories socio-professionnelles

Camille Stephanus et Josiane Vero[1]

**Proposition de résumé :**

Les salariés en emploi peu qualifié disposent d'un faible pouvoir d'agir en matière de reconversion professionnelle. Les reconversions des employés non qualifiés, pourtant fréquemment désireux de changer de métier, sont souvent empêchées ; elles s'avèrent plus fréquentes mais contraintes et externes pour les ouvriers non qualifiés.

Les reconversions professionnelles sont devenues l'un des enjeux majeurs des débats autour des politiques de l'emploi et de la formation pour résister à la concurrence mondiale et répondre aux mutations que connaît le monde du travail sous l'impulsion de dynamiques technologiques, démographiques, sociétales et économiques. D'une part, les transitions numérique et écologique suscitent des bouleversements au niveau des modes de consommation, des techniques de production, des formes d'organisation du travail, du contenu des activités. De l'autre, certains emplois peinent à trouver preneur, en raison d'un défaut d'attractivité des métiers ou de compétences des candidats.

Pour répondre à ces enjeux, la loi Avenir Professionnel du 5 septembre 2018 et le Plan d'investissement dans les compétences (PIC) ont déployé des instruments que le Plan de relance a renforcés. Les dispositifs à l'initiative des personnes deviennent les leviers d'une nouvelle responsabilisation des salariés affranchis des prescriptions de leurs employeurs. De telles orientations sont ambivalentes dans la mesure où elles oscillent entre volonté de faire plus de place à la liberté individuelle et souci de responsabiliser la personne agissante vers une adaptation à un marché du travail en crise. Ainsi, de fortes incertitudes demeurent sur marges de manœuvre dont disposent les personnes pour changer de métier.

Cette ambivalence se pose avec une acuité particulière pour les salariés en emploi peu qualifié. Ils forment un segment de main d'œuvre à part. Une plongée dans les statistiques révèle que leur volume ne faiblit pas depuis les années 1990, faisant craindre une atrophie de la classe moyenne au profit d'une polarisation de l'emploi et d'une panne de la mobilité sociale. En 2020, l'emploi non qualifié représente encore presque un salarié sur cinq. Ils sont aussi plus exposés au chômage, à la fragmentation de l'emploi et à la précarité dans un contexte où l'emploi non qualifié a changé de visage et s'est largement recomposé. L'image de l'ouvrier industriel masculin à temps complet des Trente Glorieuses s'est éclipsée au profit de la femme employée à temps partiel du secteur tertiaire (aides à domicile, caissières, assistantes maternelles, etc.), aux conditions d'emploi souvent peu favorables (faibles rémunérations, contrats courts, temps partiels, etc.) sous l'effet notamment d'une représentation syndicale et d'une culture de la négociation collective moins présente que dans les secteurs traditionnels de l'industrie. Ils sont la cible privilégiée des politiques actives de l'emploi et sont appelés à devenir les acteurs de leur vie professionnelle alors que les voies pour construire leur parcours sont loin d'être tracées.

Prendre au sérieux l'invitation qui leur est faite de se saisir de la liberté de choisir leur avenir professionnel suppose de revenir aux aspirations des personnes, d'analyser leurs mobilités concrètes entre métiers et de décrire les réalités de l'équation personnelle entre souhait de reconversion et reconversion effective. Quels sont les profils des salariés souhaitant changer de métier ? De quels métiers souhaitent-ils le plus fréquemment sortir ? Dans quelle mesure ces souhaits sont-ils concrétisés ou contrariés ? Quels types d'emplois retrouvent-ils ? Le métier retrouvé est-il proche ou au contraire très éloigné du métier de départ ? Les transitions entre métiers s'organisent-elles au sein de la même entreprise ou dans le cadre de la mobilité externe ? Conduisent-ils à des mobilités ascendantes ou descendantes ? Pour répondre à ces questions et identifier empiriquement les diverses formes de reconversion, nous analysons les souhaits et les changements de métier. Le terme « métier » s'entend au sens de la nomenclature des familles professionnelles (FAP) qui propose des niveaux de regroupements

---

de métiers plus ou moins fins en fonction de la proximité des compétences et des gestes professionnels (cf. encadré). L'étude empirique repose sur les cinq vagues d'enquêtes du dispositif Defis.

**Souhaiter changer de métier : quelles personnes et quels métiers sont concernés ?**

En 2015, selon l'enquête Defis, 33 % des salariés souhaitent changer de métier. Ce sont les employés non qualifiés qui aspirent le plus à ce changement (45%), suivis des employés qualifiés (36%) et des ouvriers non qualifiés (34%). En revanche, cadres (31%), professions intermédiaires et ouvriers qualifiés (29%) y songent moins fréquemment. A l'origine des projets de reconversions se trouvent des ressorts multiples et souvent combinés mais une composante semble omniprésente : l'insatisfaction (Stephanus et Vero, 2022). Cette insatisfaction ne surgit pas de nulle part et ne relève pas non plus d'une logique unique. Elle est de nature différente selon la catégorie socioprofessionnelle. Vouloir changer de métier illustre trois logiques repérées de façon dominante selon que l'on est salarié en emploi peu, moyennement ou très qualifié : une mise à mal de la sécurité liée à l'emploi et au salaire, un déclassement et des conditions d'emploi dégradées ou enfin une quête de sens et des aspirations plus fréquentes à laisser davantage de temps à sa vie personnelle (Ibid.). Mais quelles sont les caractéristiques des personnes et des métiers concernés ? Les résultats d'une modélisation logistique du souhait de changer de métier met en évidence les résultats suivants.

*Passé 50 ans, des souhaits de reconversion moins fréquents*

Le seuil des 50 ans est habituellement utilisé dans les études sur les travailleurs âgés. Toutes choses égales par ailleurs, leurs souhaits de reconversion sont plus faibles. Plusieurs hypothèses peuvent être formulées : le taux d'emploi des seniors, bien qu'en augmentation constante, reste faible (Insee, 2023), en particulier pour les femmes, plus souvent peu ou pas diplômées. Leurs chances de retrouver un emploi sont moins élevées attestant de freins à la demande de travail pour cette catégorie, notamment après une reconversion professionnelle (Stephanus, 2023). Au même titre que les projets portent vers l'avenir, ils sont activés par les bénéfices attendus. Or, la mobilité d'un métier à un autre devient faible à partir de 50 ans. La tendance à l'allongement de la vie professionnelle avec le recul du départ à la retraite rend pourtant cruciale la mobilité professionnelle de cette population.

*Les femmes rêvent plus souvent d'une reconversion que les hommes*

Les femmes ont une probabilité plus forte de vouloir changer de métier. Plusieurs hypothèses peuvent être avancées pour l'expliquer. D'une part, la reconversion pourrait offrir une meilleure articulation entre vie professionnelle et familiale, dont la réalisation concrète repose encore majoritairement sur des rôles sociaux genrés. D'autre part, elles ont aussi des possibilités plus limitées d'évolution et une probabilité réelle de changer de métier, supérieure à celles des hommes (Dares, 2018). La configuration familiale peut aussi jouer sur la projection dans une reconversion. Vivre en couple réduit la probabilité de souhaiter un changement de métier. En revanche, les chances de vouloir changer de métier sont modifiées au deuxième enfant et augmentent à cette occasion.

*Plus d'aspiration à changer de métier pour les employés non qualifiés*

Toutes choses égales par ailleurs, la catégorie socio-professionnelle de l'emploi occupé en 2015 intervient dans la probabilité de vouloir changer de métier. Les salariés qui nourrissent le plus fréquemment de telles aspirations occupent des postes d'employés non qualifiés. Sur les 10 métiers que les salariés souhaitent le plus fréquemment quitter pour changer de carrière, quatre concernent spécifiquement des métiers d'employés non qualifiés. Ils se répartissent dans des emplois administratifs d'entreprise, des postes de caissiers et d'employés de libre-service, des métiers de l'hôtellerie-restauration ou de l'alimentation, ou encore des emplois d'agents d'entretien et d'aides à domicile (Tableau 1). Il s'agit de métiers que l'on oppose habituellement aux emplois qualifiés dont « *l'accès* en *début* de *carrière nécessite* de *posséder* une *spécialité* de *formation spécifique* » *(*Chardon, 2002). Certains sont aujourd'hui signalés comme étant en tension par la dans leur panorama chiffré des projections à l'horizon 2030. Si la crise sanitaire a joué un rôle de déclencheur ou d'accélérateur de projets (D'Agostino et Melnik-Olive 2022), les salariés n'ont pas attendu la pandémie pour exprimer leur insatisfaction professionnelle et leurs souhaits de changement de carrière. A l'inverse, les ouvriers non qualifiés de la mécanique, du travail des métaux, du bois, des matériaux souples ou des industries graphiques envisagent moins fréquemment un changement de métier.

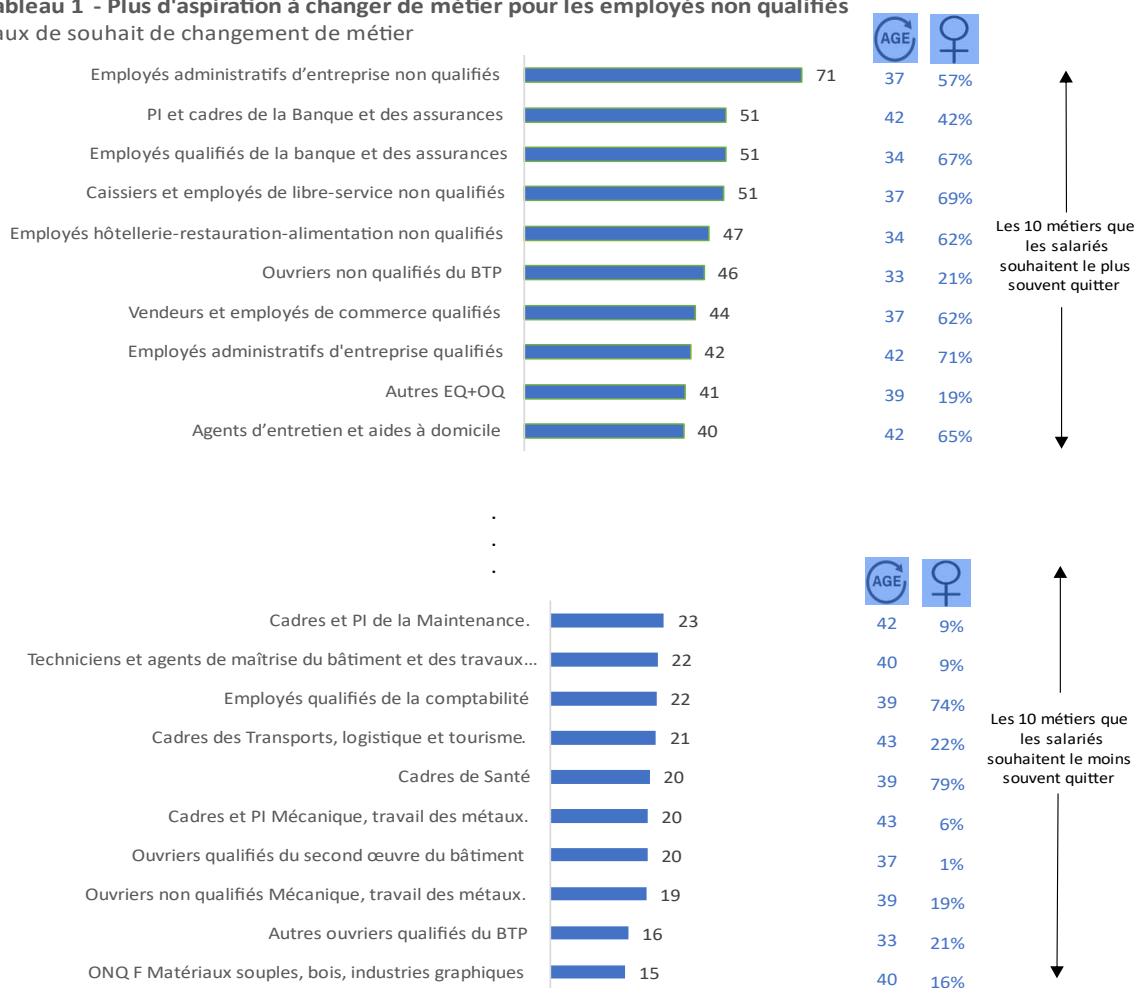

**Tableau 1 - Plus d'aspiration à changer de métier pour les employés non qualifiés**
Taux de souhait de changement de métier

*Source* : Céreq- France-Compétences-, Defis 2015
*Champ* : Salariés en décembre 2013 des entreprises du secteur privé restés dans leur entreprise jusqu'à l'été 2015
*Lecture* : 71% des employés administratifs d'entreprise souhaitent changer de métier en 2015.

## Les reconversions professionnelles au prisme des CS : du souhait à l'effectivité

La reconversion professionnelle désigne une grande variété de changements pouvant intervenir dans le parcours d'un salarié, au-delà du passage effectif d'un métier à un autre (cf. encadré). Les reconversions peuvent se faire soit en changeant d'employeur (reconversion externe), soit en restant au sein de la même entreprise (reconversion interne). Elles peuvent occasionner des changements de domaine professionnel ou opérer de légers déplacements de métiers au sein du même domaine. Elles sont susceptibles de générer des changements de statut, entre salariat et travail indépendant, et/ou des passages par le chômage, la formation ou l'inactivité. La reconversion peut aussi être promotionnelle et s'accompagner d'un passage dans la catégorie socio-professionnelle supérieure ou d'une augmentation salariale avec des responsabilités accrues, ou au contraire elle peut être qualifiée de descendante ou d'horizontale dans le cas contraire. Nous proposons dans ce qui suit un tel regard multidimensionnel sur les reconversions selon les catégories socio-professionnelles des emplois.

- Les *employés non qualifiés passent moins souvent d'un métier à l'autre.*

En 2015, ce sont les *employés non qualifiés* qui aspirent le plus à changer de métier (45%). Quatre ans plus tard, ce sont aussi eux qui sont le moins souvent passés d'un métier à un autre (32%) et dont les projets de reconversion sont plus souvent empêchés (Tableau 2).

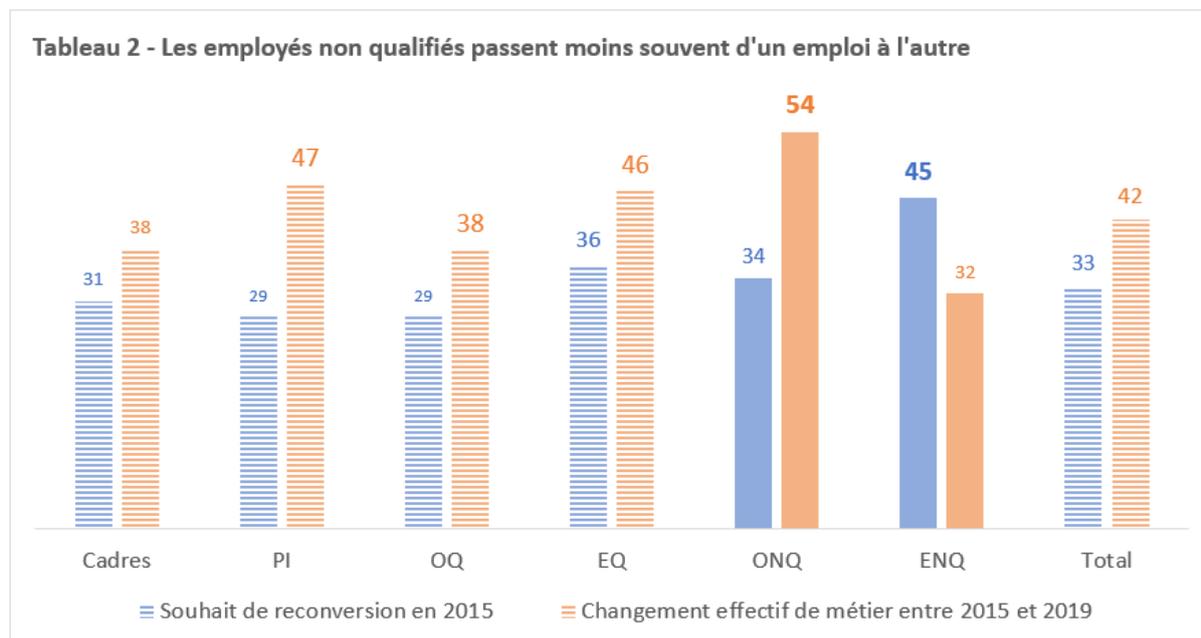

**Source** : Céreq- France-Compétences-, Defis 2015-2019
**Champ** : salariés en décembre 2013 des entreprises du secteur privé restés dans leur entreprise jusqu'à l'été 2015
**Lecture** : 45 % des employés non qualifiés souhaitent changer de métier en 2015 et seuls 32% sont passés d'un emploi à l'autre entre 2015-2019.

Une analyse synthétique des mobilités révèle que près de la moitié des employés non qualifiés (48%) restent dans le même métier chez le même employeur et n'ont donc pas connu de changement en l'espace de quatre ans malgré des souhaits plus fréquents. Elle signale aussi que 21% ont quitté ou perdu leur emploi pour un horizon professionnel incertain à quatre ans : 14 % ont connu une rupture de la relation d'emploi et sont au chômage et 7% ont changé d'emploi mais pas de métier (Tableau 3).

Néanmoins, pour ceux qui réalisent un parcours de reconversion, les changements de métier ont comme caractéristique commune de se dérouler principalement au sein même du groupe socio-professionnel des employés non qualifiés, activant ainsi une forte de circulation interne entre salariés de domaines professionnels variés : 75% des changements de métier riment avec un changement de domaine professionnel. L'analyse des données de l'enquête Defis fait aussi ressortir l'intensité des reconversions professionnelles externes et la plus grande exposition au chômage ainsi qu'à des parcours heurtés. Ces déplacements par les canaux des marchés du travail externe tracent moins souvent des parcours de mobilité promotionnelle. C'est le lot de seulement 15% des employés (Tableau 4). Les marchés externes, les déplacements de métier de grande envergure et l'absence de promotion occupent aussi une place importante dans les reconversions peu fréquentes des employés non qualifiés

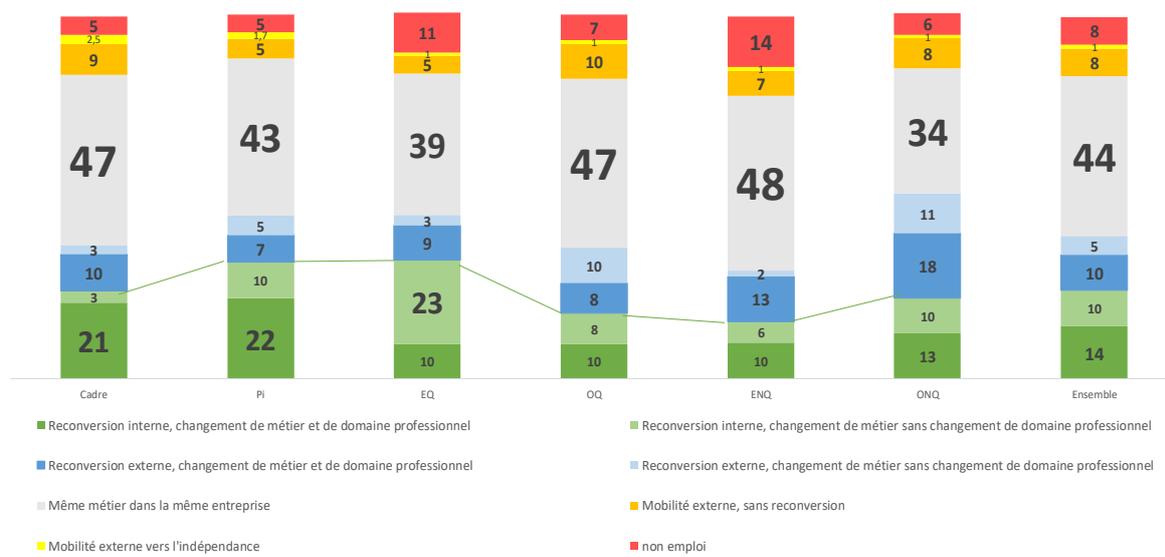

Tableau 3 – Caractérisation des mobilités entre 2015 et 2019 selon la CS du métier d'origine en 2015

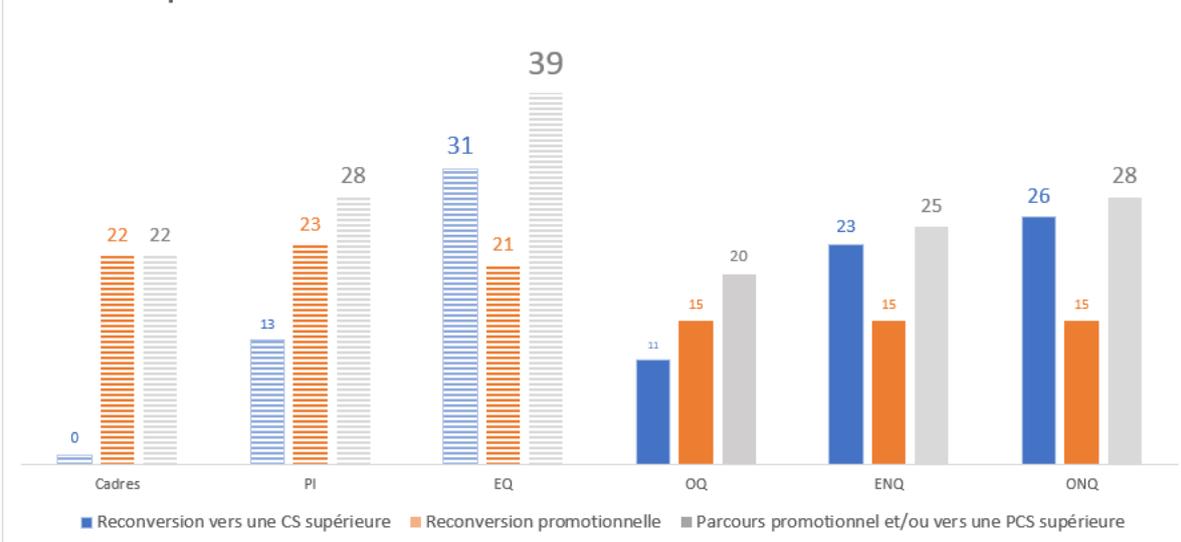

Tableau 4. Employés non qualifiés et ouvriers connaissent moins souvent une reconversion promotionnelle

*Source* : Céreq- France-Compétences-, Defis 2015-2019
*Champ* : salariés des entreprises de 10 salariés et plus du secteur privé
*Lecture* : 23% des employés non qualifiés ont changé de métier vers une CSP de niveau supérieur de qualification., 15% ont changé de métier et connu une augmentation salariale associée à une responsabilité ou une autonomie accrue. 25% d'entre eux ont changé de métier et connu l'une des deux situations précédentes.

*Intensité des reconversions externes, plus souvent externes, contraintes et rarement promotionnelles pour les ouvriers non qualifiés*

En l'espace de quatre ans, plus d'un *ouvrier non qualifié* sur deux a connu une reconversion, ce qui en fait le parcours le plus répandu de ce segment, et moins d'un tiers est resté dans la même situation professionnelle (Tableau 2). Ces chiffres illustrent l'exposition particulièrement forte des ouvriers non qualifiés aux reconversions sur la période 2015-2019 en dépit de leurs moindres souhaits exprimés en 2015 (34%, cf. Tableau 2). Cette importance des mobilités de métier se conjugue avec une plus grande exposition aux marchés externes. Ainsi, parmi les ouvriers non qualifiés passés d'un métier à l'autre, 56% ont aussi changé d'entreprise à cette occasion, contre 38% en moyenne. Ce groupe socio-professionnel constitue le volant de main d'œuvre le plus flexible (Tableau 3). Leur changement de métier se caractérise aussi par des reconversions d'envergure qui s'opèrent majoritairement dans des domaines professionnels distincts de celui d'origine. Mais cette rotation de métier n'offre pas pour autant de réelles chances de promotion. En effet, les faibles perspectives de carrière promotionnelles sont aussi un marqueur fort de leurs reconversions (Tableau 4). Aussi, pour les ouvriers non qualifiés, la logique de reconversion semble s'appuyer sur des dynamiques imprévues ou subies, majoritairement orientées vers un autre domaine professionnel, peu sécurisées, et moins souvent promotionnelles. Fréquemment suspectés de s'enfermer dans le non-emploi, des travaux soulignent en réalité qu'il n'en est rien. Ce segment de la population révèle au contraire une adaptabilité accrue aux exigences du marché du travail.

- *Les ouvriers qualifiés connaissent des reconversions modérées dans le même domaine professionnel, dans la même position sociale, mais moins souvent subies*

Loin d'être un phénomène marginal, l'exposition aux marchés externes n'est pas une particularité des salariés non qualifiés. Pour les **ouvriers qualifiés**, vouloir changer de métier se conjugue avec un changement d'employeur et des risques accrus de passage par le chômage sur la période 2015-2019. A l'instar des salariés du bas de l'échelle, changer de métier ne signifie généralement pas changer de position sociale (Tableau 4). Mais les reconversions de ce groupe professionnel restent relativement modestes et sont les plus en phase avec le niveau de souhaits exprimés en 2015.

- *Des reconversions internes et ascendantes concentrées sur les employés qualifiés, les professions intermédiaires et les cadres*

Pour les **employés qualifiés, professions intermédiaires et cadres**, et comparativement aux salariés peu qualifiés, les changements de métier sont plus en phase avec le niveau des souhaits et s'effectuent plus souvent dans la même entreprise. Si les reconversions internes offrent aux cadres de larges débouchés dans d'autres domaines professionnels, le champ des possibles se réduit pour les professions intermédiaires et les reconversions dans d'autres domaines deviennent minoritaires pour les employés qualifiés qui voient leur reconversion, certes relativement sécurisée, mais plus souvent limitée à leur domaine d'origine (Tableau 3). Un autre trait structurant des reconversions au sein de ces trois groupes sociaux professionnels concerne la plus grande fréquence de changement de métier associé à une promotion (Tableau 4).

## Conclusion

À l'heure où les politiques publiques insistent sur la liberté de choisir son avenir professionnel, en étroite association avec l'idée de responsabilité, la recherche présentée ici a souhaité examiner les formes concrètes des reconversions entre 2015 et 2019 selon le niveau de qualification des emplois.

Les résultats mettent en évidence que les salariés en emploi peu qualifié ne disposent pas du même pouvoir d'agir en matière de reconversion professionnelle que les autres salariés. Une plongée dans la dynamique des parcours montre que la structure des reconversions professionnelles des ouvriers et des employés non qualifiés diffère de celle des salariés moyennement et très qualifiés. Alors que les employés non qualifiés sont plus fréquemment désireux de changer de métier, leurs reconversions sont plus souvent empêchées. Au contraire, elles s'avèrent plus fréquentes mais contraintes pour les ouvriers non qualifiés qui multiplient les reconversions externes et connaissent les parcours professionnels les plus précaires. Néanmoins, les unes comme les autres s'avèrent plus externes et avec de faibles marges de promotion. Parallèlement, les cadres, professions intermédiaires et employés qualifiés sont aussi ceux qui concrétisent le plus souvent un parcours de reconversion, notamment promotionnel. Les aires de reconversion internes protégées leur offrent à la fois sécurité de l'emploi et perspectives de carrière, là où employés et ouvriers non qualifiés sont contraints de s'adapter aux exigences du marché du travail et indexent la sécurité de leur parcours à une adaptabilité accrue au marché du travail. De tels constats questionnent les marges de manœuvre dont disposent les salariés les moins qualifiés pour se reconvertir. Ils interrogent les moyens d'y remédier, à travers un renforcement de droits réels à des formations ambitieuses dans une logique préventive. Si chaque travailleur est appelé à devenir « acteur dans son évolution professionnelle » et à en porter la responsabilité, cela implique qu'il dispose des moyens lui permettant d'assumer une telle responsabilité. De tels moyens ne sont pas du seul ressort des salariés, mais engagent employeurs, partenaires sociaux et institutions publiques soucieuses également de répondre à la nécessaire élévation des compétences à l'heure des transitions écologiques et numériques.

## Bibliographie

Amossé T. et Chardon O. (2006), "Les travailleurs non qualifiés : une nouvelle classe sociale ?" *Économie et Statistique*, n°393-394.

Chardon O. (2002), « La qualification des employés », document de travail, n° F 0202, Insee.

D'Agostino, A., Galli, C. & Melnik-Olive, E. (2022). "Entre renoncer et se lancer : les projets de reconversion à l'épreuve de la crise », *Céreq Bref*, 427.

Dares (2005). Les familles professionnelles - Nomenclature FAP-2003, table de correspondance FAP /PCS / ROME.

Dares (2018), Changer de métier : quelles personnes et quels emplois sont concernés ? Dares Analyse n° 049, novembre 2018

Lainé F (2018). Mobilités entre les métiers et situations de travail transversales. Éclairages et Synthèses, 41.

Stephanus C., Vero J. (2022). « Se reconvertir, c'est du boulot ! Enquête sur les travailleurs non qualifiés ». Céreq Bref, 418.

Stephanus, C. (2023) « Entretien avec Camille Stephanus », *Retraite et société*, vol. 90, no. 1, 2023, pp. 167-173.

---

**Encadré : Construction des variables d'intérêt**

**La qualification de l'emploi**

En 2015, elle est déterminée à partir des déclarations des employeurs dans la base de données sociales (DADS) de 2015 classée en code PSE-ESE. Sont distingués les cadres (CS 3), des professions intermédiaires (CS 4), des ouvriers qualifiés (CS 62, 63, 64 et 65), des ouvriers non qualifiés (CS 67, 68 et 69) et des employés (CS agrégée 5). La distinction entre employés qualifiés et non qualifiés repose sur la nomenclature Chardon (2002) « *qui repose sur l'adéquation entre le contenu des emplois et la spécialité de formation des personnes qui les exercent* ». Une profession d'employé « *est ainsi définie comme qualifiée si son accès en début de carrière nécessite de posséder une spécialité de formation spécifique* » (Amossé et Chardon, 2006).

**Repérage des changements de métier**

Ici, le terme « métier » s'entend au sens de la nomenclature des familles professionnelles (FAP) qui propose des niveaux de regroupements de métiers plus ou moins fins en fonction de la proximité des compétences et des gestes professionnels (Dares 2005 ; Lainé 2018). Elle se compose de 22 domaines professionnels, déclinés en 87 familles professionnelles agrégées et 225 détaillées. Un indicateur permettant de repérer les changements de métier a été créé en s'appuyant sur une table de passage entre les codes PCS-ESE des emplois salariés issus des DADS entre 2015 et 2019 et la nomenclature des FAP en 225 familles professionnelles regroupées à un niveau plus agrégé en 22 domaines. En confrontant la FAP des individus aux différentes dates de la séquence observée, il est donc possible de distinguer ceux qui ont changé de métier des autres, au sens de la nomenclature utilisée. Un individu peut avoir changé de métier une ou plusieurs fois sur la séquence 2015-2019 ou pas du tout. Pour évaluer l'ampleur de la reconversion, on peut basiquement distinguer les métiers selon leur domaine professionnel. On considère alors qu'il y a un changement de métier plus important si l'on change de domaine professionnel. La reconversion peut aussi être promotionnelle et s'accompagner soit d'un passage dans la catégorie socio-professionnelle supérieure soit d'une augmentation du niveau de salaire associée à des responsabilités ou une autonomie accrue.

**Reconversion et mobilité de niveau de qualification de l'emploi**

La base de données DADS des années 2015 et 2019 permet d'identifier la qualification de l'emploi de chaque salarié au moment de l'enquête de 2015 et 2019, ainsi que les changements de métier (FAP) sur la période. Cela permet de distinguer les mobilités de métier caractérisées par un basculement dans un emploi de qualification supérieure ou inférieure.

**Reconversions et mobilité promotionnelles**

A chaque vague d'interrogation, l'enquête Defis recueille auprès des individus, salariés sur la période précédente, leur évolution en termes de salaires, de niveau de responsabilité et d'autonomie. Ces données permettent de construire un indicateur de mobilité promotionnelle entre deux enquêtes. La mobilité promotionnelle associe une augmentation du niveau de responsabilité ou d'autonomie à une hausse de revenus. La mobilité de CSP associe une bascule vers une CSP supérieure. Un troisième indicateur recense l'une ou l'autre des deux mobilités (mobilités promotionnelle ou de CSP).